\documentclass[10pt, a4paper]{article}
\usepackage{lrec2022} 
\usepackage{multibib}
\newcites{languageresource}{Language Resources}
\usepackage{graphicx}
\usepackage{tabularx}
\usepackage{soul}
\usepackage{titlesec}
\titleformat{\section}{\normalfont\large\bfseries\center}{\thesection.}{1em}{}
\titleformat{\subsection}{\normalfont\SmallTitleFont\bfseries\raggedright}{\thesubsection.}{1em}{}
\titleformat{\subsubsection}{\normalfont\normalsize\bfseries\raggedright}{\thesubsubsection.}{1em}{}
\renewcommand\thesection{\arabic{section}}
\renewcommand\thesubsection{\thesection.\arabic{subsection}}
\renewcommand\thesubsubsection{\thesubsection.\arabic{subsubsection}}

\usepackage{epstopdf}
\usepackage[utf8]{inputenc}

\usepackage{hyperref}
\usepackage{xstring}
\usepackage{paralist}

\usepackage{todonotes}

\usepackage{color}

\title{Multimodal Pipeline for Collection of Misinformation Data from Telegram\\ \vspace*{.5\baselineskip}} 

\name{Jose Sosa, Serge Sharoff} 

\address{University of Leeds \\
         Woodhouse Lane, LS2 9JT, Leeds, UK \\
         \{scjasm, s.sharoff\}@leeds.ac.uk\\}

\abstract{
The paper presents the outcomes of AI-COVID19, our project aimed at better understanding of misinformation flow about COVID-19 across social media platforms. The specific focus of the study reported in this paper is on collecting data from Telegram groups which are active in promotion of COVID-related misinformation. Our corpus collected so far contains around 28 million words, from almost one million messages. Given that a substantial portion of misinformation flow in social media is spread via multimodal means, such as images and video, we have also developed a mechanism for utilising such channels via producing automatic transcripts for videos and automatic classification for images into such categories as memes, screenshots of posts and other kinds of images. The accuracy of the image classification pipeline is around 87\%.
 \\ \newline  }

\begin{document}

\maketitleabstract

\section{Introduction}

The widespread distribution of COVID-19 misinformation leads to confusion, anti-health policy sentiment, and risk-tolerant behaviour \cite{chou21misinfo}. AI-COVID19 is our project which aims to use AI tools to understand the dynamics of misinformation spread across several social media platforms, including Twitter, Facebook and Telegram. This paper concentrates on the methods and outcomes of data collection from Telegram, a platform that is currently less active in policing misinformation, so that we can collect a greater variety of COVID-19 misinformation examples.

We designed a pipeline to collect COVID-19 misinformation from Telegram public channels. Then, we used this pipeline to build one of the first multimodal datasets of COVID-19 misinformation; i.e. in addition to the prevailing text data, our dataset includes images, videos, and documents. Overall the dataset comprises almost one million messages from 2k different public channels related to spreading COVID-19 misleading information. In addition, it includes 38k images, 15k videos, and 522 documents (mostly in the PDF and DOCX formats) from those channels. Furthermore, it organises the collected images into three categories: memes, posts and others by means of automatic image classification. Finally, it incorporates a set of transcripts for the collected videos.

We summarised our contributions as follows:
\begin{itemize}
    \item {Automatic pipeline for collecting misinformation from Telegram.}
    \item {Joint collection of text and multimedia data. Our pipeline for collecting data allows us to get text and also any multimedia data that can be part of the messages, e.g. images, videos, documents, audios, and stickers.}
    \item {A classifier for the collected images. We train a CNN classifier based on AlexNet \cite{krizhevsky2012imagenet} to identify memes and images from text-based posts.}
    \item {A new telegram multimodal dataset on COVID-19 related misinformation.}
 \end{itemize}

The complete code for collecting data, the classifier, and the first image of our dataset, including information about Telegram messages, users, media messages, and channels; multimedia data as classified images, videos and their transcripts will be made available for public use.\footnote{https://github.com/josesosajs/telegram-data-collection}

\section{Related work}
Since the rise of the pandemic, there have been many studies aimed at collecting resources and creating collections to deal with COVID-19 from different perspectives. Then, the number of datasets around COVID-19 significantly increased, and data sources diversified. Consequently, we can find datasets from purely scientific productions, like papers or specialised medical images, to collections of data extracted from unverified sources, like social networks.

Concerning datasets of scientific productions, the COVID-19 Open Research Dataset (CORD-19) is perhaps the best example. It contains academic articles on COVID-19 and related corona-viruses studies published between 1980 and 2021. CORD-19 represents a joint effort to provide a resource to interdisciplinary scientific communities to identify effective treatments and develop better policies for COVID-19 \cite{wang20cord}. Another reliable source of COVID-19 information is in \cite{dong2020interactive}. They developed an interactive web-based dashboard that allows real-time visualisation and tracking of reported COVID-19 cases. This practical tool served as a base for other studies in the context of COVID-19 \cite{dey2020analyzing}. 

In addition to collecting texts and statistical information, there are studies which collect COVID-related images, in particular, medical images \cite{cohen2020covid,yang2020covid,wang2021deep}. For example, \cite{xu2020deep} collected a dataset of Computer Tomography (CT) images from 110 patients with COVID-19 to train a deep neural network, which then can automatically detect the presence of COVID-19 on new CT images. Other studies rely on the collection and use of x-ray images for similar purposes \cite{hemdan2020covidx,wang2020covid,apostolopoulos2020covid}.


Automatically collecting and building datasets with those kinds of scientific information is a non-trivial process, principally because of their requirement of intense supervision to create and verify the data. However, with respect to massive amounts of unverified data, social networks represent fruitful sources to build valuable datasets around specific topics. Contrary to scientific productions, those platforms allow users to create and share any content without meticulous verification. Thus, it makes social networks a convenient medium for spreading misleading information. However, it is crucial to collect and analyse unverified data to model and understand the social response against some emerging events, e.g. COVID-19 \cite{hossain20covidlies,alam21covidmisinfo,pennycook2020fighting,brindha2020social}. 

Predominately, generic data collection from social media is performed by tracking certain accounts, posts, users, and keywords akin to the topic \cite{banda2020twitter,chen2020covid,aggarwal20reddit,basile2021dramatic,baumgartner2020pushshift}. Twitter is perhaps the most popular social network in this context. Then, we can find several COVID-19 related datasets from this platform. For instance, Banda et al. \shortcite{banda2020twitter} released a dataset of more than 150 million tweets associated with COVID-19, which represents one of the largest collections available up to date. Similar, \cite{chen2020covid} have produced a dataset of approximately 50 million tweets. Although it is a smaller dataset, it is more diverse regarding the number of languages, which makes it convenient for studies on languages other than English. Also with respect to collections in languages other than English, Alqurashi et al. \shortcite{alqurashi2020large} introduce a dataset of almost 4 million Arabic tweets linked to COVID-19. 

Analogous to Twitter, Reddit is also a valuable social media platform for building COVID-19 datasets. For example, Aggarwal et al. \shortcite{aggarwal20reddit} have produced a dataset of COVID-19 related posts and comments from Reddit, which comprises a total of 105,000 posts. Similarly, \cite{basile2021dramatic} presents an interesting collection of Reddit COVID-19 posts from different countries.  More recently, the attention has been paid other social networks, e.g. \cite{zarei2020first} collected a dataset of COVID-19 posts and comments from Instagram. Furthermore, \cite{medina20misinformation} presents a dataset which uses comments from YouTube videos to study misinformation. 

In contrast to generic data collection from social media, identifying COVID-19 misinformation is a more difficult task which requires some degree of manual verification. Some approaches managed to create misinformation datasets by a combination of data from different sources \cite{patwa2020fighting,yang2020covid,haouari2020arcov19}. \cite{cheng2021covid} created a set of annotated tweets specifically containing COVID-19 misinformation. Furthermore, some misinformation datasets cover languages like Chinese \cite{yang2021checked} and Arabic \cite{haouari2020arcov19}. Alam et al, \shortcite{alam21covidmisinfo} have produced a multilingual training set also covering the impact of misinformation, such as harmfulness and topics of their claims, for example, ``bad cure''.

Mobile messaging platforms like WhatsApp and Telegram also represent a rich source of fake and legitimate information related to the COVID-19 pandemic. The last one is the most open platform regarding access to its API. Several interesting studies in the recent literature collected and analysed Telegram data to comprehend emerging social problems like immigration movements \cite{nikkhah2018telegram}, manifestations, and terrorism \cite{prucha2016and,yayla2017telegram}. However, collecting data from Telegram is still a developing field, so there is a lack of Telegram datasets for some trending topics like COVID-19. Some studies like \cite{ng2020analyzing} collected data from this topic. However, their analysis only covered one Telegram discussion group, which does not represent the diversity of data channels. Contrarily, our approach collects data from an extensive set of Telegram public channels, which are highly related to spreading misinformation about COVID-19.  

One of the main drawbacks of many existing COVID-19 datasets from social networks like Twitter, Reddit, Youtube, and specifically Telegram is their focus on solely text data and ignoring multimedia like images and videos, which clearly impact information and misinformation flows and could also be beneficial for COVID communication research. Only few studies explored this, for example \cite{pramanick2021momenta} presents an interesting multimodal study to evaluate the harmfulness of COVID-19 oriented memes, which are abundant in most social networks. Thus, differently from other popular Telegram datasets like \cite{baumgartner2020pushshift} we additionally collected and analysed image and video data. To the best of our knowledge, our dataset is the first with respect to collecting multimodal COVID-19 data with the focus on misinformation.

\begin{figure*}[ht]
\begin{center}
\includegraphics[width=0.9\linewidth]{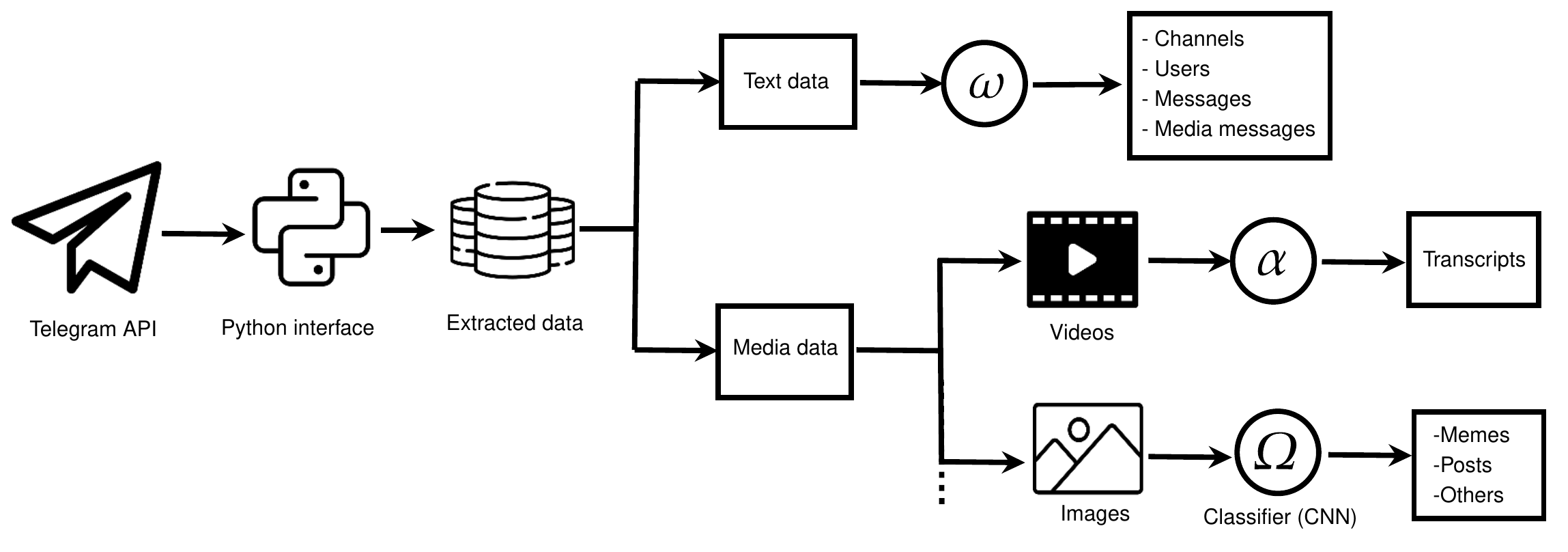} 
\caption{Telegram data collection pipeline. Our pipeline for collecting data starts with the basic block to establish the communication between our algorithm and the Telegram API. Furthermore, we designed two workflows: one for collecting texts of the messages as well as data about the channels and the users. And the second branch is for collecting media, e.g. images and videos. Then, we extended it by classifying the images with a CNN, and extracting transcripts from the videos.}
\label{fig:collection_pipeline}
\end{center}
\end{figure*}
\section{Methodology}

In this section, we describe our computational strategy for collecting data from Telegram public channels, together with our approach for analysing the media elements on some messages, e.g. images and videos.

\subsection{Collecting Data}
\label{collecting-data}

A common approach to obtaining data from messaging platforms like Telegram is by exploring the public channels. Contrary to other social networks like Twitter or Facebook, where user activity is commonly available, Telegram is not open regarding access to the messages history for specific users. Then, in line with prior approaches, our data collection from Telegram is channel-based. Similarly to \newcite{baumgartner2020pushshift} and \newcite{wich2021introducing}, we adopted a snowball sampling strategy, in which a set of seed channels leads to its augmentation by selecting channel names in messages forwarded from other channels. We started gathering data from a manually extracted list of approximately 13 public channels likely related to spreading misinformation about COVID. Then we augmented it to 70, which represented the seed for the snowball sampling strategy. Note that we manually verified that most messages from those selected channels were related to COVID misinformation.

Currently, we are collecting data from Telegram public channels daily. At the beginning of our collection process, we were retrieving messages from the 70 items in the seed list. Then, we augmented it by considering the source of forwarded messages. We repeated this process daily using the inflated collection of channels. For future iterations, due to the growing number of channels and the limitations of the Telegram API, we randomly shuffled our list and got messages from the first $n$ channels. After collecting some data, we decided to change the sorting criteria of our list of channels. Then, we ordered them by their contributions to the dataset. Thus, we assured the collection start with the $n$ elements that more messages provide to the dataset. Note that $n$ is constrained by the number of requests sent to Telegram API. On average, we are getting data from 200 public channels every day. 

\begin{figure}[ht]
\begin{center}
\includegraphics[width=\linewidth]{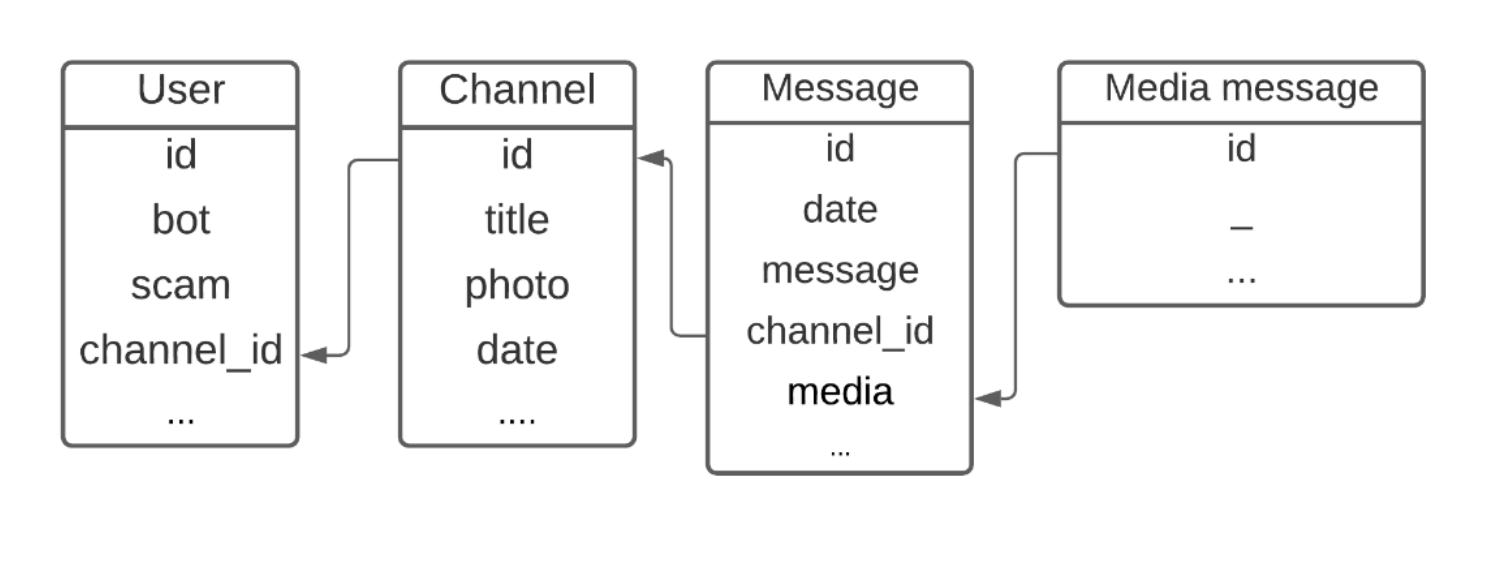} 
\caption{Structure and relationship of the JSON files for each collected Telegram entity. A complete list of the fields for each entity is include in the documentation of Telethon's API:  \url{https://tl.telethon.dev/constructors/}. The only extra field that we added is an internal UUID for each element of every entity which allows to establish the relationship between them.}
\label{fig:er-files}
\end{center}
\end{figure}

A fundamental element of our data collecting process is the interface to establish the communication between Telegram API and our algorithm. For this task,  we used Telethon\footnote{https://docs.telethon.dev/en/latest/}, a python package that allows us to configure and control the requests to the Telegram API. Then, we collected and stored data from the channels, messages and users. Furthermore, we identified and collected the messages containing media when possible. Thus, the output of our collection process is a set of four JSON files: channels, messages, users, and media messages (as illustrated on \autoref{fig:er-files}); and a folder with multiple types of media data. Having a flexible file format as the JSON files allows deploying those to any relational or non-relational database scheme.

\subsubsection{Extraction of multimodal data}
Additionally to collecting text data, we added a second branch to our pipeline to download the media from some messages, e.g. images, videos, audios, and documents. We saved those items despite their nature. However, we are particularly interested in analysing pictures and videos. 

\subsubsection{Video Data}
We collected and stored videos from a subset of telegram messages. Our early analysis of video data aims to reduce their dimensionality. Then, we extracted the transcript from each video and used the text data to summarise the video content. A transcript represents a more convenient resource for future experiments because it is less complex to analyse and classify a transcript than the video itself. For extracting the transcripts, we used $\alpha$, which represents the process of mapping from the input video $v_i$ to its respective transcript $t_i$. However this mapping does not take place directly, i.e. a function $\beta$ first extract the audio $a_i$ from $v_i$. Then, $\gamma$ inputs this intermediate representation and map it to $t_i$, which is the corresponding  transcript representation. Thus, the complete mapping is given by: 

\begin{equation}
    \alpha(v_i) : \beta(v_i) \rightarrow \gamma(a_i) \rightarrow t_i
\end{equation}

We implemented the audio extractor $\beta$ and the transcript generator $\gamma$ by means of standard python libraries \footnote{https://zulko.github.io/moviepy/ref/AudioClip.html}\footnote{ https://pypi.org/project/SpeechRecognition/}.

After having the automatically generated transcripts $t_i$, we manually evaluated their accuracy, i.e. a human annotator checked if a sample of the transcript texts matches what the videos are saying. Using the accuracy indicator, we can then classify the accurate transcripts. However, now this is still a work in progress. From a random sample of 60 transcripts, our early analysis suggests that most of those are accurate. According to the annotator 75\% of the transcripts in the subset are fairly accurate (errors are caused by such factors as background music) and around 33\% of them present COVID-19 misinformation. Then, we are expecting our future annotation follows a similar trend.

\subsubsection{Image Data}
\begin{figure}[ht]
\begin{center}
\includegraphics[width=\linewidth]{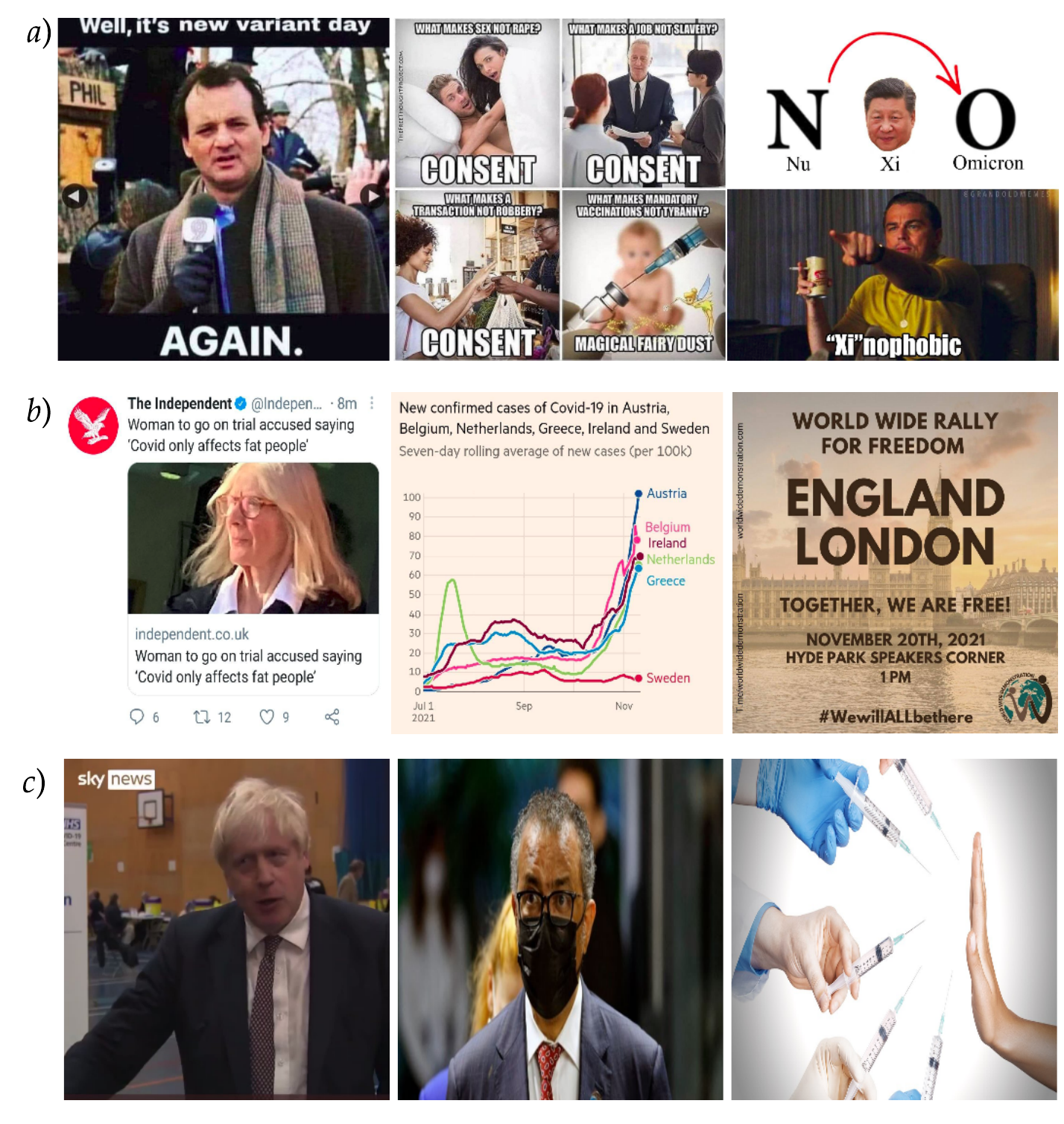} 
\caption{Random examples of classified images for each one of the three categories: $a)$ memes, $b)$ posts, and $c)$ others.
}
\label{figMemes}
\end{center}
\end{figure}

As a first approach to exploit the images, we trained a Convolutional Neural Network (CNN) $\Omega$ to distinguish between three categories: memes, posts, and others, (see \autoref{figMemes} for examples) with following definitions of the categories:

\begin{itemize}
    \item \textbf{Memes}: A meme is an image with a short piece of text, typically aimed at exciting humorous or amusing response. At this stage, the task is purely a visual classification, i.e. we are not looking for the meaning or context of the text. However, considering the nature of the channels we are following, we can assume that the collected images are highly related to COVID-19 misinformation.
    \item \textbf{Posts}: In this category, we included all the images that show posts from social media, most of them pictured as screenshots from news websites, Facebook, Twitter, and WhatsApp. Shared posters are also considered as members of this category.
    \item \textbf{Others}: This category includes all the other images that are not memes or posts. For example, we can find images of people, vaccines, masks, world leaders, pets, objects, and landscapes. 
\end{itemize}

We are particularly interested in the first two categories, i.e. memes and posts, because we can do post-processing (e.g. text extraction) and input those to existing pipelines for multimodal classification \cite{pramanick2021momenta}.

\subsubsection{Image Classifier Architecture}
\label{classifier-sec}
\begin{figure}[ht]
\begin{center}
\includegraphics[width=0.8\linewidth]{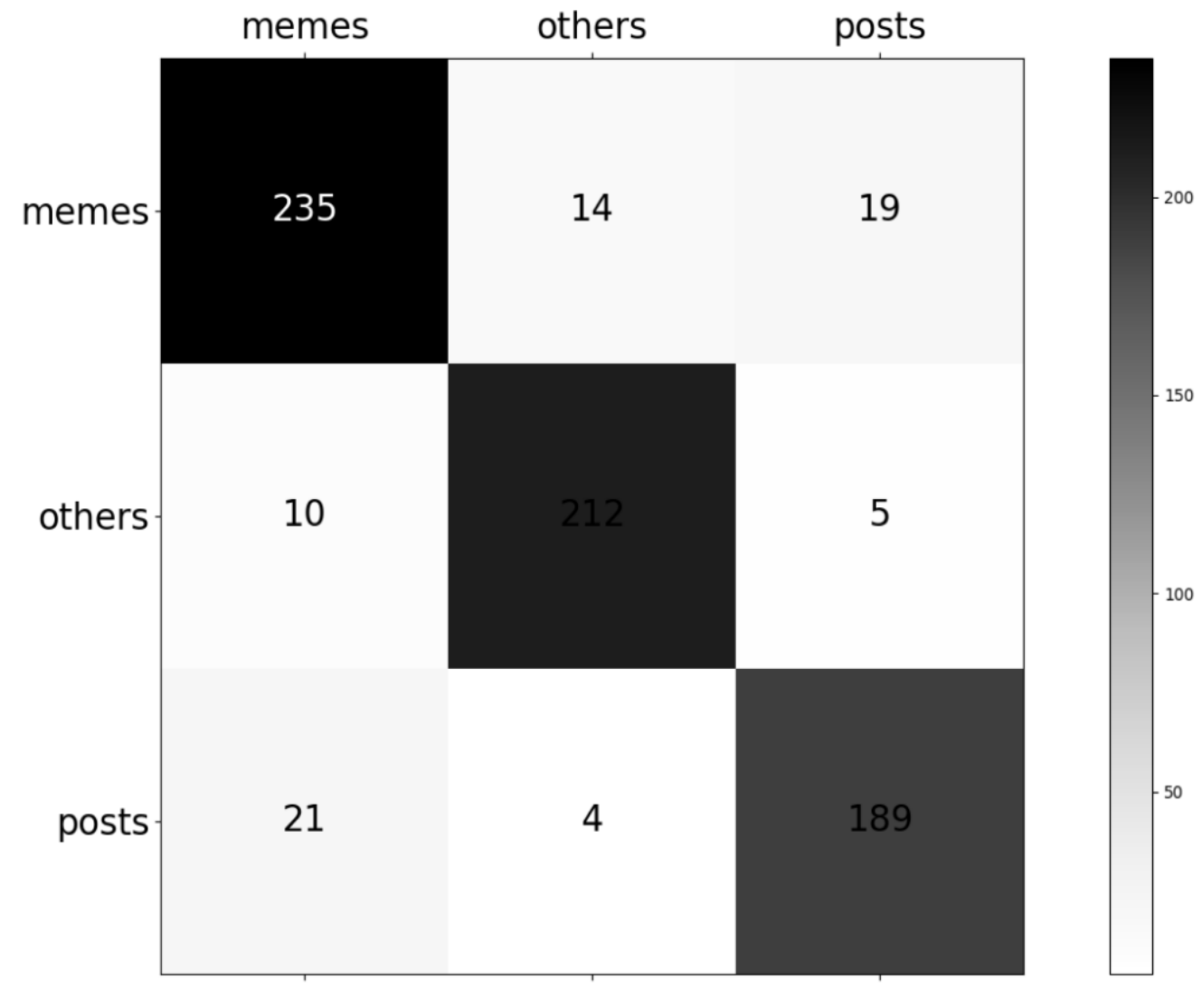} 
\caption{Confusion matrix for the three classes.}
\label{confmatrix}
\end{center}
\end{figure}

We based our CNN for image classification $\Omega$ on a pre-trained AlexNet \cite{krizhevsky2012imagenet}. We fine-tuned this model with a COVID-specific training dataset and modified the last fully connected layer to produce outputs for our three classes. For training the network $\Omega$, we utilised a subset of a publicity available dataset of Twitter images together with a dataset of COVID-specific memes\cite{singh2020mmf}. We selected approximately 3k pictures for each class and divided those in a 90/10 ratio for training and validation. Our fine-tuned classifier $\Omega$ achieved an overall accuracy of $87.4\%$ on a small test set from our Telegram collection, the confusion matrix and the scores are shown in \autoref{confmatrix} and \autoref{table:class-metrics}.

\begin{table} [ht]
\begin{center}
\begin{tabular}{ |c|c|c|c|} 
 \hline
 \textbf{Class} & \textbf{Precision} & \textbf{Recall} & \textbf{F1-score} \\ 
 \hline \hline
 Memes & 0.88 & 0.92 & 0.89\\ 
 \hline
 Others & 0.88 & 0.93 & 0.88\\ 
 \hline
 Posts & 0.88 & 0.93 & 0.89 \\ 
 \hline
\end{tabular}
\caption{Precision, Recall and F1-Score for each class.}
\label{table:class-metrics}
\end{center}
\end{table}

Similarly to the process with the videos, we extracted the text from a subset of memes and posts images $I_i$ when possible. The extraction of the text $tx_i$ from an image $I_i$ is given by $\lambda(I_i) = tx_i$, where $\lambda$ is an standard function for Optical Character Recognition (OCR)\footnote{https://github.com/madmaze/pytesseract}. However, at this stage we are not aiming for a deep analysis of the text on the memes and posts. We prioritised having an accurate classification of the images of our dataset. Then, we can use, particularly the memes as inputs for existing multimodal approaches which evaluated their harmfulness\cite{moens2021findings,zhou2021multimodal}.

\section{Results}
\begin{figure}[ht]
\begin{center}
\includegraphics[width=\linewidth]{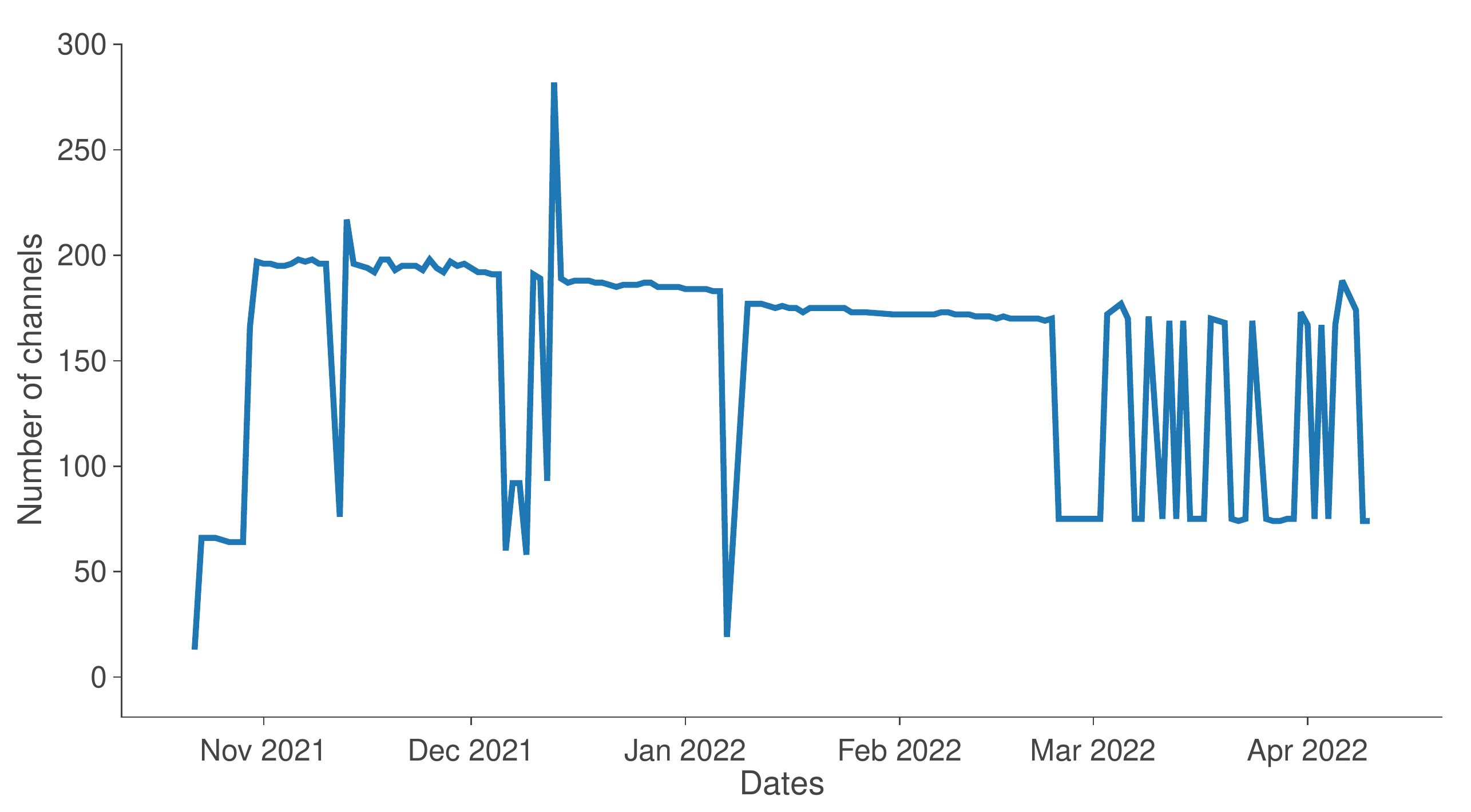} 
\caption{Number of channels extracted per day.}
\label{daily-channels}
\end{center}
\end{figure}

\textbf{Channels.} We started collecting messages from a list of approximately 13 channels highly related to contain misinformation regarding COVID-19. We augmented this list everyday as described on section \ref{collecting-data}. Then, the first version of our dataset contains messages from $2,602$ different Telegram channels. Because the limitations of Telegram API, we are not able to collect data from our full list of channels, which includes $11,161$ channels until now. Hence, after some days of collecting data, we selected a subset of channels which most contributed to our misinformation dataset, based on the number of messages and prioritise those for future data collection. This is reflected as constant line on the plot of \autoref{daily-channels}. Starting with 13 predefined channels, then 77, and from that point we increased the amount of channels automatically. In average we are collecting data from approximately 200 channels daily. There are some values below the average, those are because some unexpected issues with the collection, which normally stopped our script earlier.

\begin{figure}[ht]
\begin{center}
\includegraphics[width=\linewidth]{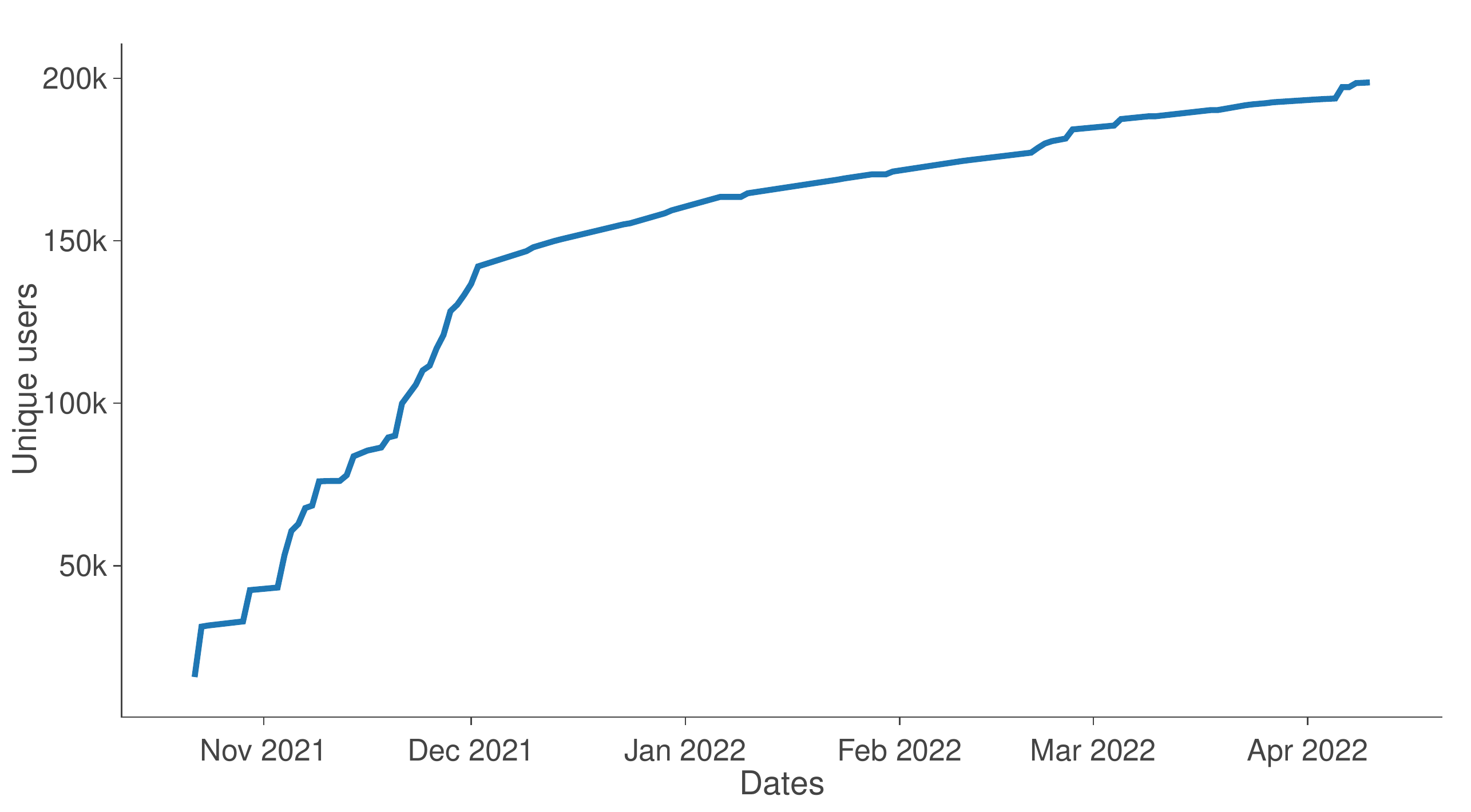} 
\caption{Unique users from all channels.}
\label{daily-users}
\end{center}
\end{figure}

\textbf{Users.} In addition to the data about the channels, we collected their users. On average our dataset contains $159,905$ unique users as shown on \autoref{daily-users}. Although the information available regarding Telegram users is quite restrictive, there are some practical indicators, e.g. if a user is a bot. Then, we can use it to determine the sort of members for a given channel.

\begin{figure}[ht]
\begin{center}
\includegraphics[width=\linewidth]{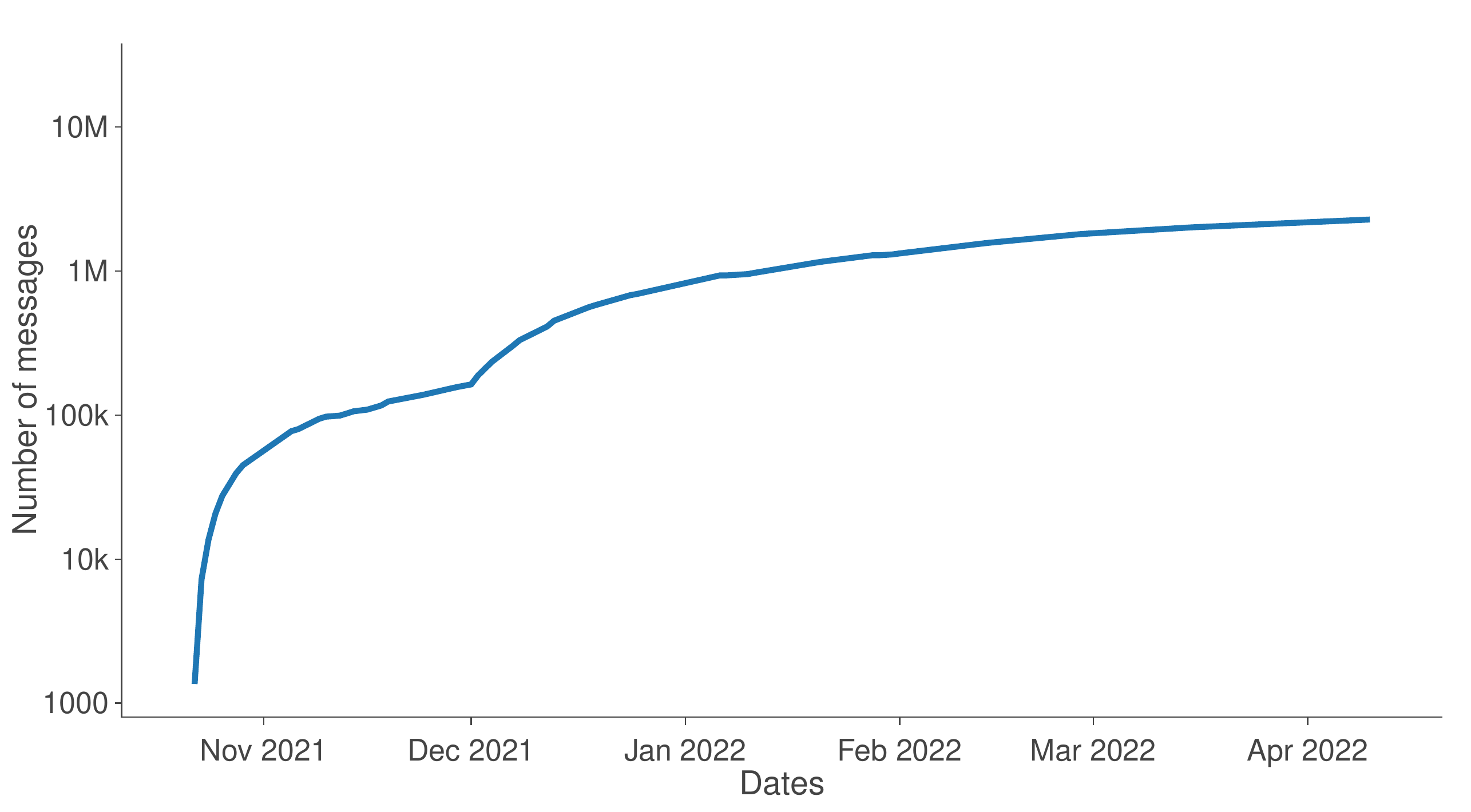} 
\caption{Accumulated telegram messages.}
\label{fig:accumulatedmessages}
\end{center}
\end{figure}

\textbf{Messages.} Overall the first snapshot of our dataset contains $1,131,560$. After ignoring the empty messages, which represent service messages from Telegram or messages with only media, we got a total of $812,196$ messages that have some text (See \autoref{fig:accumulatedmessages}). We started to collect data from October $22^{nd}$, its first release version includes data until December $31^{st}$ of 2021, with the graphs reported in this paper updated until April 2022. As our data collection is still running, we expect our dataset to continue growing.

We used a language classifier \cite{lui2012langid}\footnote{https://github.com/saffsd/langid.py} for automatic language identification for each message. Although English is the predominant language in our pipeline, our collection results contain a significant proportion of messages in other languages as illustrated in Table ~\ref{table:lang-dist}. In average the length of messages in English is $256$ characters, which is quite long, i.e. almost twice the length of text data obtained from other social networks like Twitter. 

\begin{table}[ht]
\begin{center}
\begin{tabular}{ |l|c|c|} 
 \hline
 \textbf{Language} & \textbf{Messages($\downarrow$)} & \textbf{Percent} \\ 
 \hline \hline
 English (en) & 549,126 & 67.61 \\ 
 \hline
 German (de) & 58,308 & 7.18 \\ 
 \hline
 Chinese (zh)  & 35,051 & 4.32\\
 \hline
 Spanish (es)& 24,821 & 3.06\\
 \hline
 Russian (ru) & 21,098 & 2.60\\
 \hline
 Others & 106, 630 & 15.23\\
 \hline
\end{tabular}
\caption{Languages distribution of all the messages in our dataset.}
\label{table:lang-dist}
\end{center}
\end{table}

Additionally, we calculated the frequency for each word in our corpus, only including English Telegram messages. Then, we used the frequency list of OpenWebTex\cite{sharoff2020know} as a reference corpus for computing the log-likelihood (LL). This metric helps to identify the most indicative words in our corpus when compared against the reference. \autoref{tab:ll-message} shows the top 20 words sorted by their log-likelihood score. 

\begin{table}[ht]
    \centering
    \small
    \begin{tabular}{|l |l |l| l|} 
        \hline
        \textbf{Word}   & \textbf{F1} & \textbf{F2} & \textbf{LL$(\downarrow)$}\\
        \hline \hline
        vaccine&132959&26026&179230\\
        vaccinated&26075& 9086&71786\\
        vaccines&90899& 9471&54248\\
        vaccination&48375& 7312&46913\\
        unvaccinated& 7366& 4063&35175\\
        jab&21719& 4708&33285\\
        pandemic&19085& 3646&24948\\
        vax& 2222& 2501&24268\\
        coronavirus& 1204& 2097&21489\\
        mandates&32631& 3339&19013\\
        virus&192947& 5061&16019\\
        jabbed& 2345& 1620&14590\\
        jabs& 8847& 2005&14335\\
        booster&35224& 2669&13700\\
        chronology&11605& 2016&13447\\
        adverse&75885& 3223&13057\\
        ivermectin&  810& 1207&12145\\
        vaxxed& 1068& 1184&11465\\
        deaths&331266& 4854&10387\\
        passports&32779& 2007& 9501\\
        \hline
    \end{tabular}
    \caption{Frequencies and Log-likelihood scores (LL) of representative words appearing on the Telegram messages.}
    \label{tab:ll-message}
\end{table}

\textbf{Hashtags and mentions.} Furthermore, following previous works \cite{baumgartner2020pushshift}, we looked for hashtags and mentions within the messages, obtaining a list of $9,127$ unique hashtags, from a set of $53,728$ elements and $127,248$ mentions, from which $5,468$ are unique mentions. We observed some evident hashtags from the list on \autoref{table:hashtags} directly related to COVID-19. For example, \#Omicron, \#COVID19, \#Vaccines, and \#Covid. For the case of mentions in Table \ref{table:mentions}, most of them correspond to channels dedicated to shared news, predominately fake ones. However, two channels of the list got the Telegram verified badge; @disclosetv and @EpochTimes, which are neutral respecting their shared content.

To verify our snowball strategy, we assessed random 50 messages from each of the original set of 13 channels, the 70 channel extended set and from our current collection. The rate of misinformation messages drops from 74\% to 60\% to 42\%,  while still keeping the majority of messages in our target category.

\begin{table} [ht]
\begin{center}
\small
\begin{tabular}{ |l|c|} 
 \hline
 \textbf{Hashtag} & \textbf{Proportion(\%)}\\ 
 \hline \hline
    \#KAG & 12.84  \\
    \#WeAreTheNewMedia & 5.43  \\
    \#Omicron & 1.00  \\
    \#COVID19 & 0.98  \\
    \#WWG1WGA & 0.84  \\
    \#ShutItDown & 0.66  \\
    \#FightBack & 0.49  \\
    \#CrimesAgainstHumanity & 0.39  \\
    \#UnitedWeStand & 0.36  \\
    \#MAGA & 0.34  \\
    \#ReclaimTheLine & 0.32  \\
    \#China & 0.31  \\
    \#DoNotComply & 0.30  \\
    \#SaveTheChildren & 0.30  \\
    \#Ukraine & 0.30  \\
    \#Vaccines & 0.29  \\
    \#TheDefender & 0.29  \\
    \#UndergroundWarReport & 0.28  \\
    \#Covid & 0.25  \\
    \#Antifa & 0.24  \\
    \hline
\end{tabular}
\caption{Most common hashtags from our corpus of Telegram messages. Note that we performed this analysis exclusively on English messages.}
\label{table:hashtags}
\end{center}
\end{table}

\begin{table} [ht]
\begin{center}
\small
\begin{tabular}{ |l|c|} 
 \hline
 \textbf{Mention} & \textbf{Prop.(\%)}\\ 
 \hline \hline
    @WeTheNews & 5.66  \\
    @PookztA & 5.42  \\
    @PatriotArmy & 5.42  \\
    @SergeantRobertHorton & 4.26  \\
    @disclosetv & 4.11  \\
    @ZeroHedgeTyler & 2.28  \\
    @AreWeAllBeingPlayed & 2.09  \\
    @EpochTimes & 1.73  \\
    @HoCoMDPatriots & 1.66  \\
    @KanekoaTheGreat & 1.40  \\
    @leagueofextraordinarypepes & 1.38  \\
    @TGNewsU & 1.35  \\
    @No\_BS\_News & 1.23  \\
    @OneRepublicNetwork & 1.21  \\
    @HATSTRUTH & 1.20  \\
    @ChiefNerd & 1.07  \\
    @CBKNEWS & 1.07  \\
    @ExposeThePEDOSendTheCABAL & 0.98  \\
    @awakenspecies & 0.87  \\
    @GitmoTV & 0.84  \\
    \hline
\end{tabular}
\caption{Most common mentions from our corpus of Telegram messages. Similar to the hashtags, we only used English messages.}
\label{table:mentions}
\end{center}
\end{table}

\subsection{Multimodal data}
\begin{figure}[ht]
\begin{center}
\includegraphics[width=\linewidth]{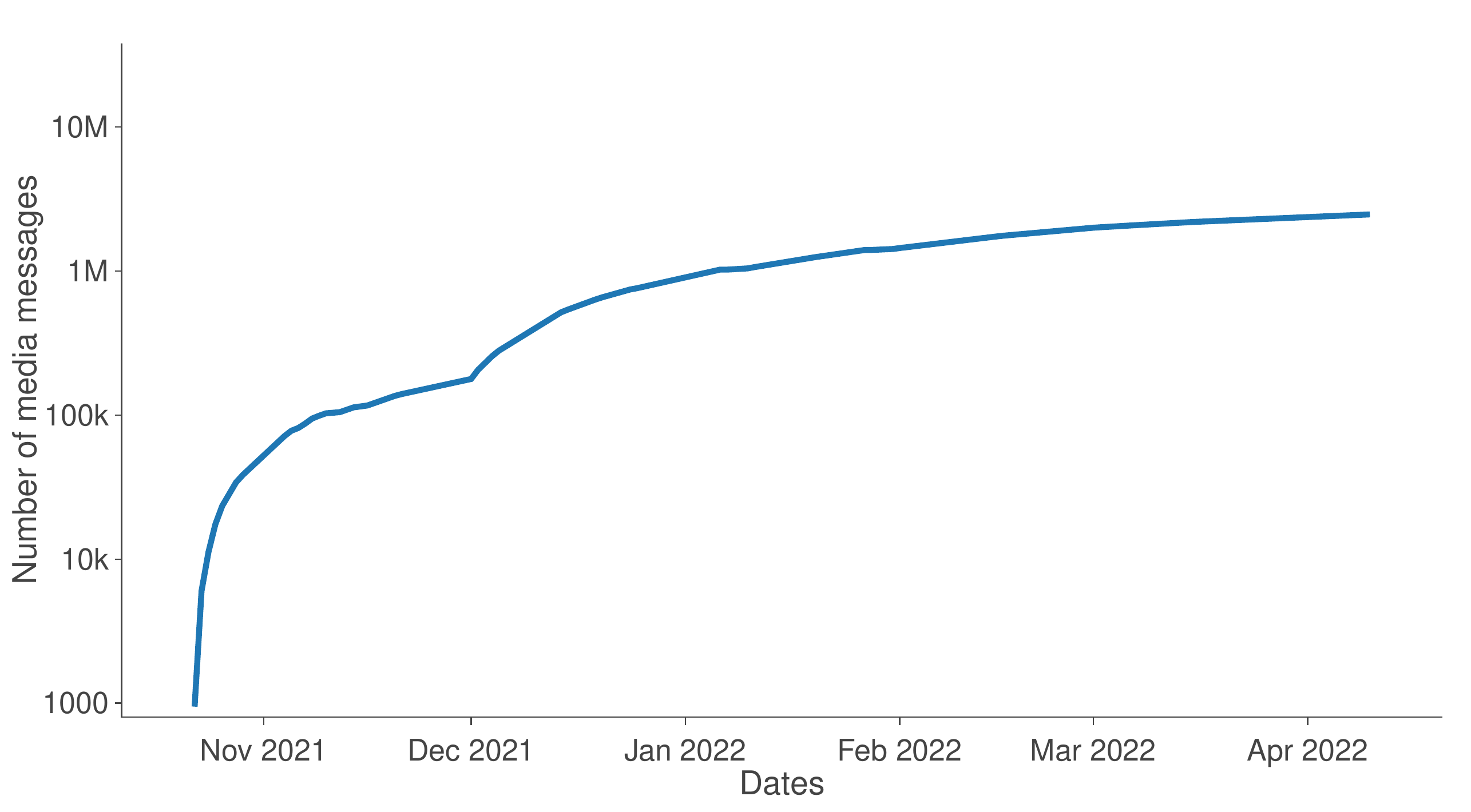} 
\caption{Accumulated media messages.}
\label{fig:medi_messages}
\end{center}
\end{figure}

\textbf{Media messages.} From a total of $1,131,560$ Telegram messages, $888,810$ include some media data, either accompanying the text or solely the media itself (See Figure \ref{fig:medi_messages}). Because the limitations of the number of request send to the Telegram API, we are not able to download all the media from those messages. However, we built a JSON file for saving them and keep the reference to the file, which we can use those to download it if still available in the future.

\begin{figure}[ht]
\begin{center}
\includegraphics[width=0.9\linewidth]{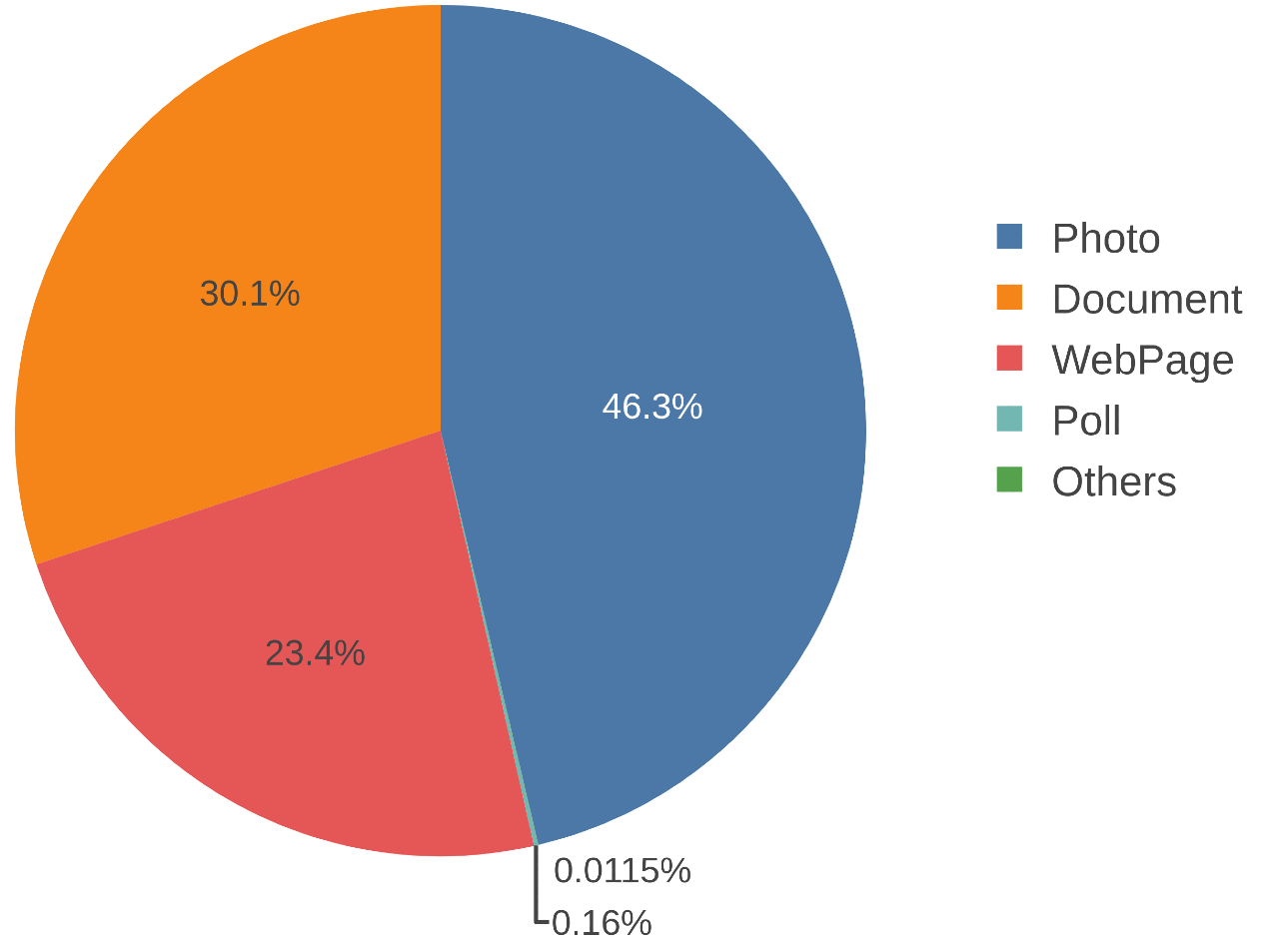} 
\caption{Media messages types distribution. Others includes: Invoice, Unsupported, Game, Dice, and Contact.}
\label{media-types-distribution}
\end{center}
\end{figure}

The media messages are distributed into nine categories determined by Telegram: \textit{Photo, Document, WebPage, Poll, Invoice, Unsupported, Game, Dice,} and \textit{Contact}. According to Figure \ref{media-types-distribution}, \textit{Photo}, \textit{Document} and \textit{WebPage} are the dominant categories on our dataset. Note that Telegram includes videos under the category of \textit{Document}.

\begin{figure}[ht]
\begin{center}
\includegraphics[width=\linewidth]{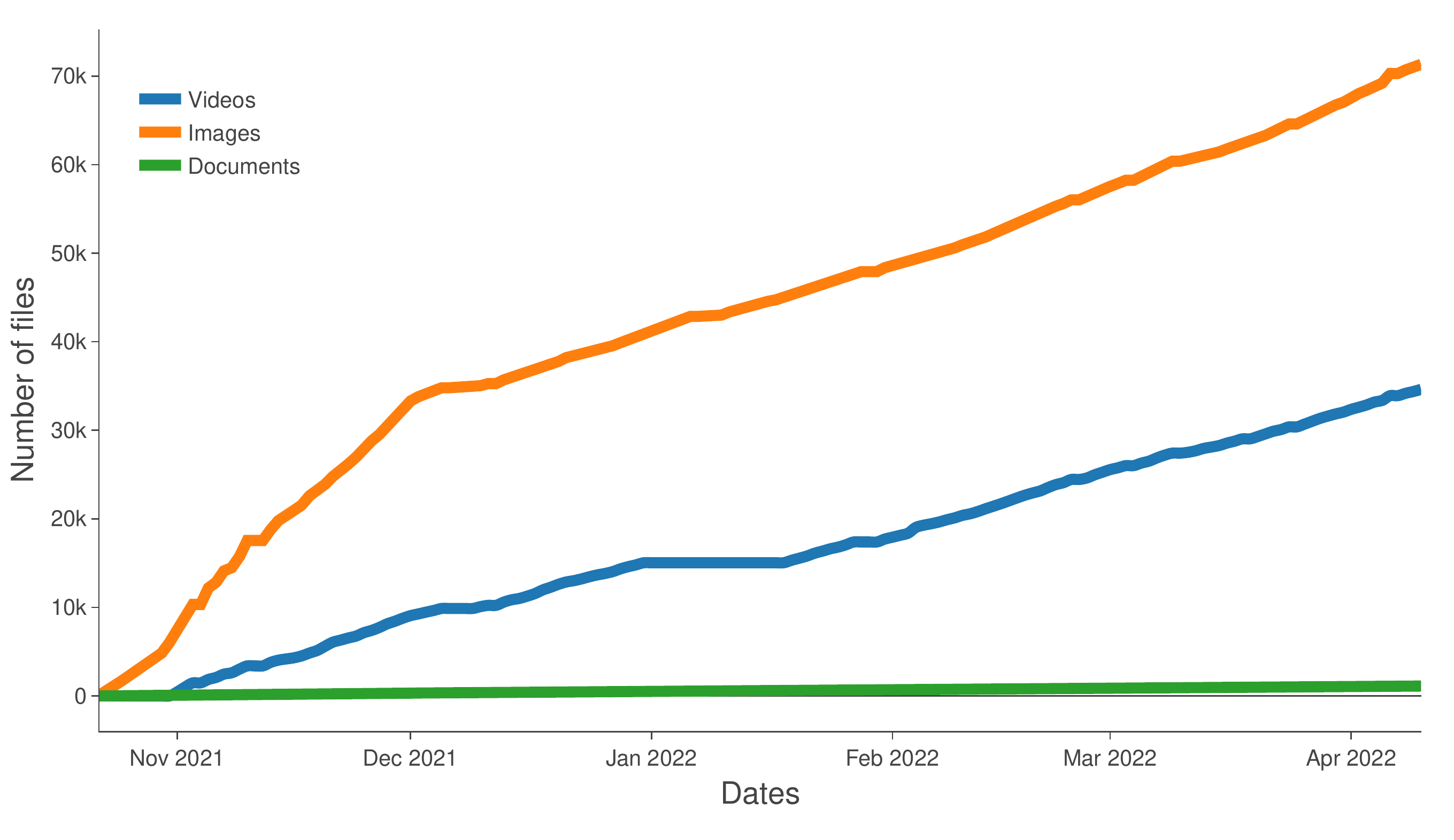} 
\caption{Accumulated media. Daily distribution of collected images, videos, and documents.}
\label{fig:accumaltedmediamultimodal}
\end{center}
\end{figure}

We downloaded the files from media messages when possible despite their category, which extended our dataset. Currently, it comprises $40,882$ images, $15,040$ videos, and $522$ documents (.pdf, .doc, etc). \autoref{fig:accumaltedmediamultimodal} shows the number of collected files every day.  At this stage, we are particularly interested in the analysis of videos and images.

\begin{figure}[ht]
\begin{center}
\includegraphics[width=\linewidth]{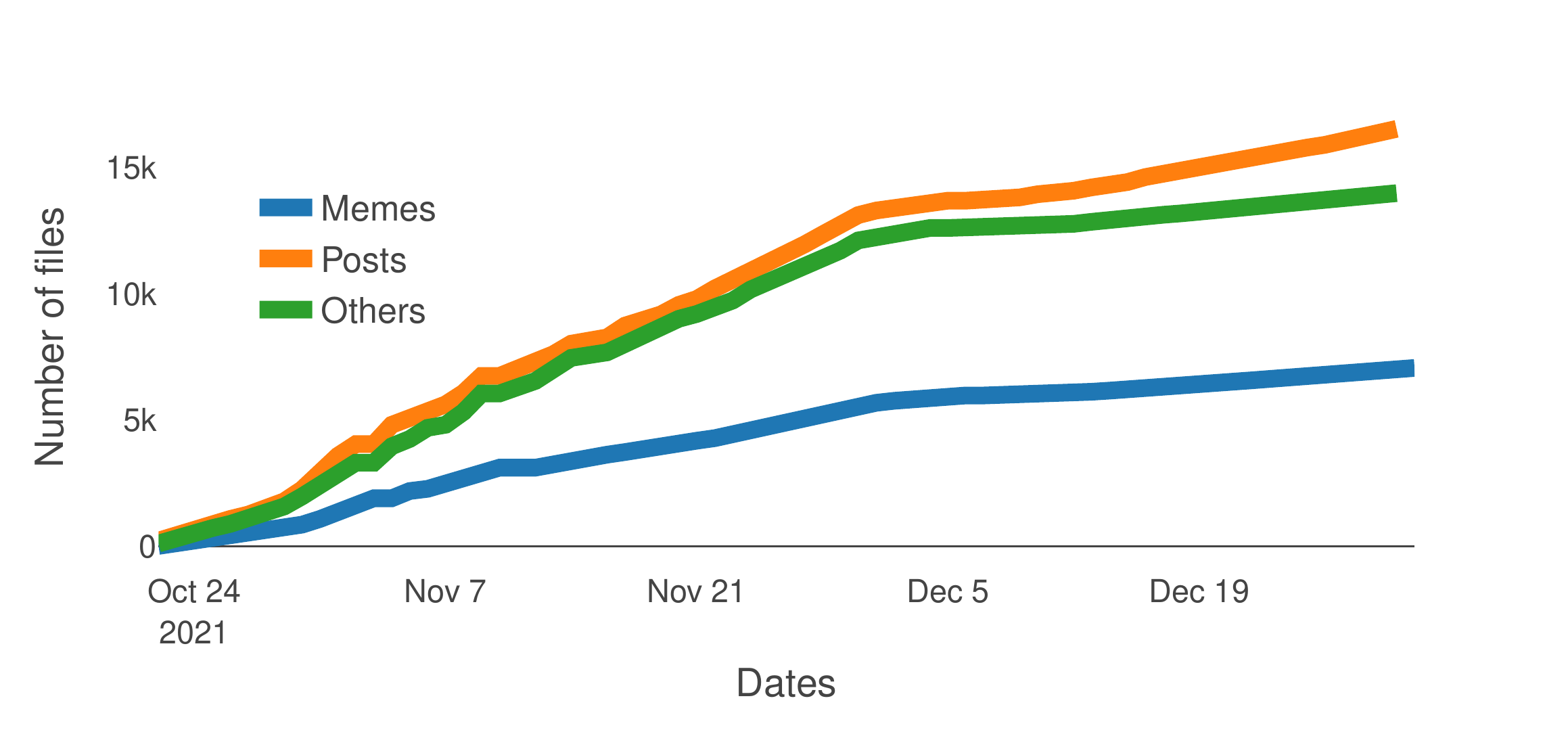} 
\caption{Distribution of classified images.}
\label{fig:classified_images}
\end{center}
\end{figure}

\textbf{Images.} After removing duplicates from the whole set of images, we used our trained classifier described in section \ref{classifier-sec} to classify the remainder $37, 616$ into three classes: Memes, posts, and others. As shown in Figure \ref{fig:classified_images} the amount posts (16,546) and others (13,999) are quite similar, while the number of memes (7,071) represents almost half of them.

\textbf{Videos and transcripts.} We extracted the transcripts from the collected videos and built a corpus of $3,231,184$ words from them. Then, we calculated the log-likelihood score, using the same process as for the messages. Table \ref{table:freq-words} lists the top 30 words from this corpus, based on their log-likelihood score. We only removed function words\footnote{https://www.nltk.org/}. Note that most of the words in the list are related to COVID-19,  which is a positive indicator to assume the context of the videos is in the COVID19 domain without explicitly looking at them.

\begin{table} [ht]
\begin{center}
\begin{tabular}{|l |l |l| l|} 
 \hline
    \textbf{Word}   & \textbf{F1} & \textbf{F2} & \textbf{LL($\downarrow$)}\\
    \hline\hline
    vaccine&132959& 5943&44861\\
    vaccinated&26075& 2210&19408\\
    people&15749678&19211&18879\\
    vaccines&90899& 2759&18751\\
    virus&192947& 3157&17664\\
    f******& 2482& 1324&15992\\
    coronavirus& 1204& 1063&13653\\
    pandemic&19085& 1083& 8676\\
    vaccination&48375& 1277& 8332\\
    pfizer&12224&  748& 6101\\
    booster&35224&  865& 5522\\
    5g&40781&  852& 5171\\
    ivermectin&  810&  404& 4836\\
    myocarditis&  384&  364& 4712\\
    graphene&34240&  748& 4605\\
    immune&146846& 1099& 4499\\
    lockdown&19436&  628& 4344\\
    oxide&27241&  660& 4196\\
    omicron&  706&  344& 4105\\
    adverse&75885&  843& 4086\\
    children&2504217& 3483& 4075\\
    pcr&16345&  506& 3458\\
    flu&64332&  704& 3392\\
    protein&213869& 1015& 3291\\
    viruses&63719&  687& 3291\\
    hydroxychloroquine&  122&  237& 3283\\
    unvaccinated& 7366&  411& 3279\\
    disease&574392& 1502& 3275\\
    vaccinations&18834&  481& 3107\\
    cells&521885& 1395& 3095\\
    \hline

\end{tabular}
\caption{Log-likelihood score (LL) of words appearing on the transcripts. }
\label{table:freq-words}
\end{center}
\end{table}

\section{Conclusions and Future Work} 

In this paper we described our pipeline for collecting multimodal data from Telegram. Futhermore we detailed the first version of our misinformation dataset, to the best of our knowledge it is the first Telegram dataset to include multimodal misinformation data about COVID-19. Our dataset includes almost one million Telegram messages from approximately 2K channels. Additionally, it comprises around 60k multimedia files, distributed between images, videos and documents. Furthermore, we report an automatic classifier for the image categories, and a transcript extraction tool for the collected videos. 

Our dataset represents a valuable resource for researchers from different disciplines. For example, our collection of memes could augment similar existing datasets for multimodal analysis. Similarly, our corpus of messages and video transcripts could serve to study the flow of COVID-19 misinformation in social networks. This early version of the dataset represent one of the most complete and structured Telegram collections around COVID-19 misinformation. We plan to continue collecting data in the same way and expand our current dataset within the next months. Furthermore, we will perform an extensive analysis of the images, videos and transcripts using multimodal analysis frameworks \cite{knight2020multimodal}.


\section{Acknowledgements}

Project has been funded via the EPSRC IAA3125 grant: AI tracing tools for detecting COVID19 misinformation in collaboration with the Yorkshire\&Humber Academic Health Sciences Network and the British National Health Service.

\section{Bibliographical References}
\label{main:ref}
\bibliographystyle{lrec2022-bib}
\bibliography{lrec2022-example}

\begin{thebibliography}{}

\bibitem[\protect\citename{Aggarwal \bgroup et al.\egroup
  }2020]{aggarwal20reddit}
Aggarwal, J., Rabinovich, E., and Stevenson, S.
\newblock (2020).
\newblock Exploration of gender differences in {COVID-19} discourse on
  {R}eddit.
\newblock In {\em Proceedings of the 1st Workshop on {NLP} for {COVID-19} at
  {ACL} 2020}, Online, July. Association for Computational Linguistics.

\bibitem[\protect\citename{Alam \bgroup et al.\egroup
  }2021]{alam21covidmisinfo}
Alam, F., Shaar, S., Dalvi, F., Sajjad, H., Nikolov, A., Mubarak, H.,
  Da~San~Martino, G., Abdelali, A., Durrani, N., Darwish, K., Al-Homaid, A.,
  Zaghouani, W., Caselli, T., Danoe, G., Stolk, F., Bruntink, B., and Nakov, P.
\newblock (2021).
\newblock Fighting the {COVID-19} infodemic: Modeling the perspective of
  journalists, fact-checkers, social media platforms, policy makers, and the
  society.
\newblock In {\em Findings of the Association for Computational Linguistics:
  EMNLP 2021}, pages 611--649, Punta Cana, Dominican Republic, November.
  Association for Computational Linguistics.

\bibitem[\protect\citename{Alqurashi \bgroup et al.\egroup
  }2020]{alqurashi2020large}
Alqurashi, S., Alhindi, A., and Alanazi, E.
\newblock (2020).
\newblock Large arabic twitter dataset on {COVID-19}.
\newblock {\em arXiv preprint arXiv:2004.04315}.

\bibitem[\protect\citename{Apostolopoulos and
  Mpesiana}2020]{apostolopoulos2020covid}
Apostolopoulos, I.~D. and Mpesiana, T.~A.
\newblock (2020).
\newblock {COVID-19}: automatic detection from x-ray images utilizing transfer
  learning with convolutional neural networks.
\newblock {\em Physical and Engineering Sciences in Medicine}, 43(2):635--640.

\bibitem[\protect\citename{Banda \bgroup et al.\egroup }2020]{banda2020twitter}
Banda, J.~M., Tekumalla, R., Wang, G., Yu, J., Liu, T., Ding, Y., and Chowell,
  G.
\newblock (2020).
\newblock A twitter dataset of 150+ million tweets related to {COVID-19} for
  open research.
\newblock {\em Type: dataset}.

\bibitem[\protect\citename{Basile \bgroup et al.\egroup
  }2021]{basile2021dramatic}
Basile, V., Cauteruccio, F., and Terracina, G.
\newblock (2021).
\newblock How dramatic events can affect emotionality in social posting: The
  impact of {COVID-19} on reddit.
\newblock {\em Future Internet}, 13(2):29.

\bibitem[\protect\citename{Baumgartner \bgroup et al.\egroup
  }2020]{baumgartner2020pushshift}
Baumgartner, J., Zannettou, S., Squire, M., and Blackburn, J.
\newblock (2020).
\newblock The pushshift telegram dataset.
\newblock In {\em Proceedings of the International AAAI Conference on Web and
  Social Media}, volume~14, pages 840--847.

\bibitem[\protect\citename{Brindha \bgroup et al.\egroup
  }2020]{brindha2020social}
Brindha, M.~D., Jayaseelan, R., and Kadeswara, S.
\newblock (2020).
\newblock Social media reigned by information or misinformation about
  {COVID-19}: a phenomenological study.
\newblock {\em Alochana Chakra Journal}, 9(5):585--602.

\bibitem[\protect\citename{Chen \bgroup et al.\egroup }2020]{chen2020covid}
Chen, E., Lerman, K., and Ferrara, E.
\newblock (2020).
\newblock {COVID-19}: The first public coronavirus twitter dataset.
\newblock {\em Type: dataset}.

\bibitem[\protect\citename{Cheng \bgroup et al.\egroup }2021]{cheng2021covid}
Cheng, M., Wang, S., Yan, X., Yang, T., Wang, W., Huang, Z., Xiao, X.,
  Nazarian, S., and Bogdan, P.
\newblock (2021).
\newblock A {COVID-19} rumor dataset.
\newblock {\em Frontiers in Psychology}, 12.

\bibitem[\protect\citename{Chou \bgroup et al.\egroup }2021]{chou21misinfo}
Chou, W.-Y.~S., Gaysynsky, A., and Vanderpool, R.~C.
\newblock (2021).
\newblock The {COVID-19} misinfodemic: Moving beyond fact-checking.
\newblock {\em Health Education \& Behavior}, 48(1):9--13.

\bibitem[\protect\citename{Cohen \bgroup et al.\egroup }2020]{cohen2020covid}
Cohen, J.~P., Morrison, P., Dao, L., Roth, K., Duong, T.~Q., and Ghassemi, M.
\newblock (2020).
\newblock {COVID-19} image data collection: Prospective predictions are the
  future.
\newblock {\em arXiv preprint arXiv:2006.11988}.

\bibitem[\protect\citename{Dey \bgroup et al.\egroup }2020]{dey2020analyzing}
Dey, S.~K., Rahman, M.~M., Siddiqi, U.~R., and Howlader, A.
\newblock (2020).
\newblock Analyzing the epidemiological outbreak of {COVID-19}: A visual
  exploratory data analysis approach.
\newblock {\em Journal of medical virology}, 92(6):632--638.

\bibitem[\protect\citename{Dong \bgroup et al.\egroup
  }2020]{dong2020interactive}
Dong, E., Du, H., and Gardner, L.
\newblock (2020).
\newblock An interactive web-based dashboard to track {COVID-19} in real time.
\newblock {\em The Lancet infectious diseases}, 20(5):533--534.

\bibitem[\protect\citename{Haouari \bgroup et al.\egroup
  }2020]{haouari2020arcov19}
Haouari, F., Hasanain, M., Suwaileh, R., and Elsayed, T.
\newblock (2020).
\newblock Arcov19-rumors: Arabic {COVID-19} twitter dataset for misinformation
  detection.
\newblock {\em arXiv preprint arXiv:2010.08768}.

\bibitem[\protect\citename{Hemdan \bgroup et al.\egroup
  }2020]{hemdan2020covidx}
Hemdan, E. E.-D., Shouman, M.~A., and Karar, M.~E.
\newblock (2020).
\newblock Covidx-net: A framework of deep learning classifiers to diagnose
  {COVID-19} in x-ray images.
\newblock {\em arXiv preprint arXiv:2003.11055}.

\bibitem[\protect\citename{Hossain \bgroup et al.\egroup
  }2020]{hossain20covidlies}
Hossain, T., Logan~IV, R.~L., Ugarte, A., Matsubara, Y., Young, S., and Singh,
  S.
\newblock (2020).
\newblock {COVIDL}ies: Detecting {COVID-19} misinformation on social media.
\newblock In {\em Proceedings of the 1st Workshop on {NLP} for {COVID-19} (Part
  2) at {EMNLP} 2020}, Online, December. Association for Computational
  Linguistics.

\bibitem[\protect\citename{Knight and Adolphs}2020]{knight2020multimodal}
Knight, D. and Adolphs, S.
\newblock (2020).
\newblock Multimodal corpora.
\newblock In {\em A practical handbook of corpus linguistics}, pages 353--371.
  Springer.

\bibitem[\protect\citename{Krizhevsky \bgroup et al.\egroup
  }2012]{krizhevsky2012imagenet}
Krizhevsky, A., Sutskever, I., and Hinton, G.~E.
\newblock (2012).
\newblock Imagenet classification with deep convolutional neural networks.
\newblock {\em Advances in neural information processing systems},
  25:1097--1105.

\bibitem[\protect\citename{Lui and Baldwin}2012]{lui2012langid}
Lui, M. and Baldwin, T.
\newblock (2012).
\newblock langid.py: An off-the-shelf language identification tool.
\newblock In {\em Proceedings of the {ACL} 2012 System Demonstrations}, pages
  25--30, Jeju Island, Korea, July. Association for Computational Linguistics.

\bibitem[\protect\citename{Medina~Serrano \bgroup et al.\egroup
  }2020]{medina20misinformation}
Medina~Serrano, J.~C., Papakyriakopoulos, O., and Hegelich, S.
\newblock (2020).
\newblock {NLP}-based feature extraction for the detection of {COVID-19}
  misinformation videos on {Y}ou{T}ube.
\newblock In {\em Proceedings of the 1st Workshop on {NLP} for {COVID-19} at
  {ACL} 2020}, Online, July. Association for Computational Linguistics.

\bibitem[\protect\citename{Moens \bgroup et al.\egroup
  }2021]{moens2021findings}
Moens, M.~F., Huang, X.-J., Specia, L., and Yih, W.-t.
\newblock (2021).
\newblock Findings of the association for computational linguistics: Emnlp
  2021.
\newblock In {\em Findings of the Association for Computational Linguistics:
  EMNLP 2021}.

\bibitem[\protect\citename{Ng and Loke}2020]{ng2020analyzing}
Ng, L. H.~X. and Loke, J.~Y.
\newblock (2020).
\newblock Analyzing public opinion and misinformation in a {COVID-19} telegram
  group chat.
\newblock {\em IEEE Internet Computing}, 25(2):84--91.

\bibitem[\protect\citename{Nikkhah \bgroup et al.\egroup
  }2018]{nikkhah2018telegram}
Nikkhah, S., Miller, A.~D., and Young, A.~L.
\newblock (2018).
\newblock Telegram as an immigration management tool.
\newblock In {\em Companion of the 2018 ACM Conference on Computer Supported
  Cooperative Work and Social Computing}, pages 345--348.

\bibitem[\protect\citename{Patwa \bgroup et al.\egroup
  }2020]{patwa2020fighting}
Patwa, P., Sharma, S., Pykl, S., Guptha, V., Kumari, G., Akhtar, M.~S., Ekbal,
  A., Das, A., and Chakraborty, T.
\newblock (2020).
\newblock Fighting an infodemic: {COVID-19} fake news dataset.
\newblock {\em arXiv preprint arXiv:2011.03327}.

\bibitem[\protect\citename{Pennycook \bgroup et al.\egroup
  }2020]{pennycook2020fighting}
Pennycook, G., McPhetres, J., Zhang, Y., Lu, J.~G., and Rand, D.~G.
\newblock (2020).
\newblock Fighting {COVID-19} misinformation on social media: Experimental
  evidence for a scalable accuracy-nudge intervention.
\newblock {\em Psychological science}, 31(7):770--780.

\bibitem[\protect\citename{Pramanick \bgroup et al.\egroup
  }2021]{pramanick2021momenta}
Pramanick, S., Sharma, S., Dimitrov, D., Akhtar, M.~S., Nakov, P., and
  Chakraborty, T.
\newblock (2021).
\newblock Momenta: A multimodal framework for detecting harmful memes and their
  targets.
\newblock {\em arXiv preprint arXiv:2109.05184}.

\bibitem[\protect\citename{Prucha}2016]{prucha2016and}
Prucha, N.
\newblock (2016).
\newblock Is and the jihadist information highway--projecting influence and
  religious identity via telegram.
\newblock {\em Perspectives on Terrorism}, 10(6):48--58.

\bibitem[\protect\citename{Sharoff}2020]{sharoff2020know}
Sharoff, S.
\newblock (2020).
\newblock Know thy corpus! robust methods for digital curation of web corpora.
\newblock {\em arXiv preprint arXiv:2003.06389}.

\bibitem[\protect\citename{Singh \bgroup et al.\egroup }2020]{singh2020mmf}
Singh, A., Goswami, V., Natarajan, V., Jiang, Y., Chen, X., Shah, M., Rohrbach,
  M., Batra, D., and Parikh, D.
\newblock (2020).
\newblock Mmf: A multimodal framework for vision and language research.
\newblock \url{https://github.com/facebookresearch/mmf}.

\bibitem[\protect\citename{Wang \bgroup et al.\egroup }2020a]{wang2020covid}
Wang, L., Lin, Z.~Q., and Wong, A.
\newblock (2020a).
\newblock Covid-net: A tailored deep convolutional neural network design for
  detection of {COVID-19} cases from chest x-ray images.
\newblock {\em Scientific Reports}, 10(1):1--12.

\bibitem[\protect\citename{Wang \bgroup et al.\egroup }2020b]{wang20cord}
Wang, L.~L., Lo, K., Chandrasekhar, Y., Reas, R., Yang, J., Burdick, D., Eide,
  D., Funk, K., Katsis, Y., Kinney, R.~M., Li, Y., Liu, Z., Merrill, W.,
  Mooney, P., Murdick, D.~A., Rishi, D., Sheehan, J., Shen, Z., Stilson, B.,
  Wade, A.~D., Wang, K., Wang, N. X.~R., Wilhelm, C., Xie, B., Raymond, D.~M.,
  Weld, D.~S., Etzioni, O., and Kohlmeier, S.
\newblock (2020b).
\newblock {CORD-19}: The {COVID-19} open research dataset.
\newblock In {\em Proceedings of the 1st Workshop on {NLP} for {COVID-19} at
  {ACL} 2020}, Online, July. Association for Computational Linguistics.

\bibitem[\protect\citename{Wang \bgroup et al.\egroup }2021]{wang2021deep}
Wang, S., Kang, B., Ma, J., Zeng, X., Xiao, M., Guo, J., Cai, M., Yang, J., Li,
  Y., Meng, X., et~al.
\newblock (2021).
\newblock A deep learning algorithm using ct images to screen for corona virus
  disease ({COVID-19}).
\newblock {\em European radiology}, pages 1--9.

\bibitem[\protect\citename{Wich \bgroup et al.\egroup
  }2021]{wich2021introducing}
Wich, M., Gorniak, A., Eder, T., Bartmann, D., Cakici, B.~E., and Groh, G.
\newblock (2021).
\newblock Introducing an abusive language classification framework for telegram
  to investigate the german hater community.
\newblock {\em arXiv preprint arXiv:2109.07346}.

\bibitem[\protect\citename{Xu \bgroup et al.\egroup }2020]{xu2020deep}
Xu, X., Jiang, X., Ma, C., Du, P., Li, X., Lv, S., Yu, L., Ni, Q., Chen, Y.,
  Su, J., et~al.
\newblock (2020).
\newblock A deep learning system to screen novel coronavirus disease 2019
  pneumonia.
\newblock {\em Engineering}, 6(10):1122--1129.

\bibitem[\protect\citename{Yang \bgroup et al.\egroup }2020]{yang2020covid}
Yang, X., He, X., Zhao, J., Zhang, Y., Zhang, S., and Xie, P.
\newblock (2020).
\newblock Covid-ct-dataset: a ct scan dataset about {COVID-19}.
\newblock {\em arXiv preprint arXiv:2003.13865}.

\bibitem[\protect\citename{Yang \bgroup et al.\egroup }2021]{yang2021checked}
Yang, C., Zhou, X., and Zafarani, R.
\newblock (2021).
\newblock Checked: Chinese {COVID-19} fake news dataset.
\newblock {\em Social Network Analysis and Mining}, 11(1):1--8.

\bibitem[\protect\citename{Yayla and Speckhard}2017]{yayla2017telegram}
Yayla, A.~S. and Speckhard, A.
\newblock (2017).
\newblock Telegram: The mighty application that isis loves.
\newblock {\em International Center for the Study of Violent Extremism}, 9.

\bibitem[\protect\citename{Zarei \bgroup et al.\egroup }2020]{zarei2020first}
Zarei, K., Farahbakhsh, R., Crespi, N., and Tyson, G.
\newblock (2020).
\newblock A first instagram dataset on {COVID-19}.
\newblock {\em arXiv preprint arXiv:2004.12226}.

\bibitem[\protect\citename{Zhou \bgroup et al.\egroup
  }2021]{zhou2021multimodal}
Zhou, Y., Chen, Z., and Yang, H.
\newblock (2021).
\newblock Multimodal learning for hateful memes detection.
\newblock In {\em 2021 IEEE International Conference on Multimedia \& Expo
  Workshops (ICMEW)}, pages 1--6. IEEE.

\end{thebibliography}

\end{document}